 \documentclass[final,5p,times,twocolumn]{elsarticle}

\usepackage{amssymb}
\usepackage{lineno}
\usepackage[colorlinks=true,
            linkcolor=red,
            urlcolor=blue,
            citecolor=gray]{hyperref}
\usepackage{graphicx}	
\usepackage{amsmath}	
\usepackage{amssymb}	
\usepackage{subfig}
\usepackage{cleveref}
\usepackage{nameref}
\usepackage{array,multirow,makecell}
\usepackage{tabularx}
\usepackage[dvipsnames]{color}
\usepackage{float}
\usepackage{longtable} 

\journal{Astroparticle Physics}

\begin{document}

\begin{frontmatter}



\title{A Single Zone Synchrotron Model for Flares of PKS1510-089}


\author{Partha Pratim Basumallick}
\ead{basuparth314@gmail.com}
\author{Nayantara Gupta}


\address{Raman Research Institute,
    C.V.Raman Avenue, Sadashivanagar, Bangalore 560080, India}

\begin{abstract}
PKS 1510-089 is one of the most variable blazars. Very high energy gamma ray emission from this source was observed by H.E.S.S. during March-April 2009 and by MAGIC from February 3 to April 3, 2012 quasi-simultaneously with multi-wavelength flares. The spectral energy distributions of these flares have been modeled earlier with the external Compton mechanism which depends on our knowledge of the densities of the seed photons in the broad line region, the dusty infrared torus or a hypothetical slow sheath surrounding the jet around the radio core. Here we show that to explain the multi-wavelength data with synchrotron emission of electrons and protons the jet power should be of the order of $10^{48}$ ergs/sec.  
\end{abstract}


\end{frontmatter}


\section{Introduction}
With the successful operation of the high energy gamma ray detectors it  has become possible to study blazar flares even at very high energies. 
Blazars have relativistic jets directed toward us where the radiation losses from the relativistic electrons and protons result in the emission of photons of radio to gamma ray frequencies. 
\par They are powerful sources of GeV-TeV gamma rays.  Their spectral attenuations at very high energy with cosmological distances are widely used to study the absorption of high energy gamma rays by the extragalactic background light (EBL) \cite{DK}.
\par Their spectral energy distributions (SEDs) have  two broad peaks. The peak at low frequency originates from synchrotron radiations of accelerated electrons in the relativistic jet and the peak at higher frequency from the inverse Compton scattering of these electrons by the seed photons in the jet, and the external regions including the disc torus, the broad line region (BLR).
 \par The high energy hump may also originate from synchrotron emission of accelerated protons in the jet and hadronic interactions \cite{Bo09,Diltz}. Hadronic interactions could be between accelerated protons and cold matter or low energy photons. 
 
\par Flat Spectrum Radio Quasars (FSRQs) and BL Lacs are two different classes of blazars having different spectral features. FSRQs have bright optical and UV emission lines \cite{Massaro}, also 
they are more luminous in high energy photons than BL Lacs.  External Compton (EC) mechanism successfully describes the high energy emission from FSRQs in most cases while synchrotron self Compton (SSC) emission is the most popular scenario for BL Lacs. 
 \par PKS 1510-089, PKS 1222+21, and 3C 279 are the three FSRQs observed in gamma rays of energy more than 100 GeV. The flares from 3C 279 have been modeled earlier with the synchrotron emission from accelerated electrons and protons, also with the two zone SSC model \cite{Paliya}.
 
 \par PKS 1510-089 is a FSRQ located at a redshift of 0.361 with highly polarised radio and optical emission. Although many FSRQs have been found to have a spectral break in the frequency range of a few GeV \cite{Abdo11}, the SED of PKS 1510-089 shows no such distinctive feature. In fact the transition from the high energy (HE:100 MeV to 100 GeV) to VHE range is quite smooth for the observed data.
 
\par It has a black hole of mass $5.6\times 10^8$ times the mass of sun estimated from its accretion disc temperature and UV flux \cite{Abdo10}.
 Fermi LAT and AGILE detectors have detected highly variable gamma ray emission from it. Detection of gamma rays of energy  upto 300---400 GeV has been reported by the H.E.S.S. collaboration \cite{Abram} during March-April, 2009 and the MAGIC collaboration \cite{Alek}  from February 3 to April 3, 2012.
 
\par The quasi-simultaneous data from radio, optical, X-ray and gamma ray telescopes have been combined and modeled with EC mechanism in \cite{BK}. They showed that the H.E.S.S. data given in \cite{H.E.S.S.13}  could be included in their single zone model  if  EC emission happens due to the seed photons in the BLR and the dusty torus of PKS 1510-089. They also included the effect of the internal absorption of the gamma rays due to pair production with the photons in the blazar environment. Their model is primarily based on the assumption that the emission region is located near the central core i.e., the near dissipation zone scenario \cite{Abdo11}. Moreover, the absorption of the gamma rays by the EBL further attenuates the gamma ray spectrum before detection. 
 
 \par However according to the findings of \cite{Marscher} it is possible that the emission region is located at large distances (tens of parsecs) from the blazar core. This would mean that the BLR cloud and the IR dusty torus can no longer act as possible sources of seed photons. In order to overcome this complication it has been hypothesized that the jet may have components moving with different relative velocities which provide the necessary seed photons for IC scattering. The authors of \cite{Alek} have considered both the far and near dissipation zone scenarios while modeling the SED obtained from observations. For the near dissipation zone scenario the IR torus was considered as the source of seed photons. Whereas for the far dissipation zone scenario they have considered a slow moving sheath enveloping the faster moving spine of the jet as the source of seed photons.
  
 \par PKS 1510-089 was detected in flaring states by the HESS and MAGIC telescopes in 2009 and 2012 respectively. In the present work we have used a single zone lepto-hadronic model to explain the multi-wavelength data during these flares. The lower energy peak is attributed to electron synchrotron mechanism whereas we consider the proton synchrotron mechanism for explaining the higher energy bump. Unlike the previous models we do not need to consider external sources of photons to explain the observed spectra. The following sections describe our work in detail.
\section{Modeling the SED}
\label{section_2}
The multi-wavelength data presented in \cite{BK} and \cite{Alek} [Observation timeline :-- March--April(2009) $\&$ February--April(2012) respectively] have been modeled in our work. PKS 1510-089  was active in the VHE regime during these  periods.  
 \par We have used the code developed in \cite{Krawczynski} to generate the synchrotron spectra from relativistic electrons and protons in the jet of  PKS 1510-089.
To model the higher energy bump in the SED, we consider a population of relativistic protons in this energy range $10^{15}<E^{proton}<10^{20}$ eV.
 The relativistic protons are losing energy by synchrotron emission at a much slower rate compared to the relativistic electrons. 
 Diffusion losses are  assumed to be negligible for the relativistic protons in our model. Also,  the energy loss of  the very high energy protons due to $p \, \gamma$ interactions is found to be insignificant compared to synchrotron losses for the parameter values used in our study.
 Their injected  spectrum remains essentially unchanged as their cooling time scales are much larger than the time scale of the flares in our study. The injected spectrum ($\frac{dQ(E^{proton})}{dE^{proton}}$) of protons is assumed to follow a broken power law  with spectral indices $p_1$ and $p_2$ below and above the break energy respectively.  In our model $p_1\sim 2$ and  the values of $p_2$, the break energy are adjusted to fit the observed SED. Due to inefficient cooling of protons we need very high luminosity in protons to explain the high energy data in the SEDs of flares with proton synchrotron emission.
  \par
 In our model the most important cooling mechanism for relativistic electrons is synchrotron emission.  Their injected spectrum is assumed to be a simple power law with $p_1\sim 2$.  
 The emission region is assumed to be a spherical blob of radius $R$ moving with relativistic speed.
 The break energy in the propagated spectrum of electrons is calculated using the following condition
\begin{equation}
t^{electron}_{synch}\simeq\frac{7.75\times 10^{8}}{B^{2}\times \gamma^{electron}_{break}}=\frac{R}{c} 
\label{eqn_1}  
\end{equation}
where $\gamma^{electron}_{break}=\frac{E^{electron}_{break}}{m_{electron}c^{2}}$.
Beyond the break energy the propagated spectrum of electrons steepens by a factor of $1/E_{electron}$ due to synchrotron cooling. We note that the light crossing time ($R/c$) is comparable to the time scale of the flares measured in the jet frame. The above condition ensures that below the cooling break energy the synchrotron photons from the entire spherical region of radius $R$ contribute to the observed flux.
 
  The radius of the spherical emission region (\textit{R} in cm) is larger than the Larmor radius  of the highest energy protons ($E^{proton}_{max}$  in eV ) in our model. Eqn--\ref{eqn_2} represents this condition below 
  
\begin{equation}
B\geqslant 30 \frac{E^{proton}_{max}}{10^{19}} \frac{10^{15}}{R} \qquad \rm  {in\  Gauss}.
\label{eqn_2}
\end{equation}
The ambient magnetic field (\textit{B}) is constrained by eqn (\ref{eqn_1}) and eqn (\ref{eqn_2}). 
We have chosen two different values for \textit{B}, 0.42 G \& 0.62 G  while modeling the observed SEDs. 

 \par The parameters such as redshift (z = 0.361 for PKS 1510-089), bulk Lorentz factor($\Gamma$), viewing angle($\theta_{obs}$) are dependent on direct observations. Although we cannot measure $\Gamma$ and $\theta_{obs}$ in the same way we can measure the redshift, the observed apparent velocity ($\sim$ 20\textit{c}---45\textit{c}) of the relativistic jet \cite{Jorstad} does indicate a very small viewing angle (1.4$^{\circ}$
--- 3$^{\circ}$) \cite{Homan01,Homan02,Marscher,Cabrera} which in turn constrains the value of the Lorentz factor according to the relation $\Gamma \simeq \frac{1}{\theta_{obs}}$. Due to the extremely high apparent superluminal motion displayed by the jet the value of the Doppler factor $\left(\delta = \frac{1}{\Gamma (1-\beta \cos \theta)}\right)$ should also be very high. We have selected $\delta$= 20 \& 36 by appropriately adjusting the values of $\theta_{obs}$ $\&$ $\Gamma$ while fitting the observed data. The code from \cite{Krawczynski} used in our work  also requires the energy density of the particles in the jet frame $u_{particle}^{'}$ as a input parameter. 
  As the jet power is directly proportional to the energy density of particles, a lower value of this parameter is always preferred.

  \par The radius of the emission region is generally constrained by the flux variability of the source according to the relation $R{\leq}\frac{c\Delta t_{obs}\delta}{1+z}$ where $\Delta t_{obs}$ is the observed variability timescale. As no significant variability was detected during the VHE emissions either by H.E.S.S. or MAGIC we cannot infer any strong upper limit on the radius.  However in the HE $\gamma$-ray regime AGILE-GRID has registered 7 days and 14 days for two distinctive flares, Flare-I and Flare-II respectively in 2012 \cite{Alek}. Also, the HE data shows variability in shorter time scale which may originate from fluctuations in magnetic field and particle density within a smaller region. For $\delta = 36$, the radius of the emission region for the AGILE-GRID observations should be \(4.79 {\times} 10^{17}\) and \(9.58 {\times} 10^{17}\) cms for Flare-I and II respectively. Similarly for $\delta = 20$ the estimates of \textit{R} are found to be \(2.66 {\times} 10^{17}\) and \(5.33 {\times} 10^{17}\) cms respectively for Flare-I and Flare-II. The variability in the radio frequency range was also in the timescale of weeks during that observing period. Also during the observations in 2009 \cite{BK} two distinct flares were detected by \textit{Fermi}-LAT within a duration of 42 days. It is important to mention in this context that the above relation is an approximate one and might lead to large errors while estimating the dimensions of the emitting region \cite{Protheroe}. Thus considering all the factors discussed above we have assumed the value of \textit{R} as \(5.7{\times}10^{17}\) cm for our emission region which is in agreement with the choice of \textit{R} used by \cite{BK} \& \cite{Alek} previously while modeling the 2009 \& 2012 observations respectively. Although this choice of \textit{R} is not consistent with the hour scale variability during the $\gamma$-ray emission reported by \textit{Fermi}-LAT (\cite{Brown},\cite{Foschini},\cite{Saito}) it must be noted that as the multiwavelength data reported during the two high activity states of PKS 1510-089 (2009 and 2012) are at best quasi-simultaneous in nature  the SED represents an average emission state of the source. As a result of this the models presented in our paper and also  in \cite{BK}, \cite{Alek} are not expected to account for the separately observed shorter timescale variabilities. According to \cite{Marscher}, \cite{Marscher14} \& \cite{Narayan} the rapid flickering may be attributed to turbulence within the jet flow and it need not be absolutely essential that the volume of the emission region should have these rapid variabilities as a constraining factor. It may be possible that isolated regions with slightly different magnetic field signatures and particle populations with a different energy range and energy density exist within the larger emitting region and the rapid flickering may originate in these regions. Future studies along the line of \cite{Marscher14} is expected to facilitate our understanding of this intricate emission mechanism.       
        \par
\begin{figure*}
\centering
  \begin{tabular}{@{}ccc@{}}
    \includegraphics[width=.48\textwidth]{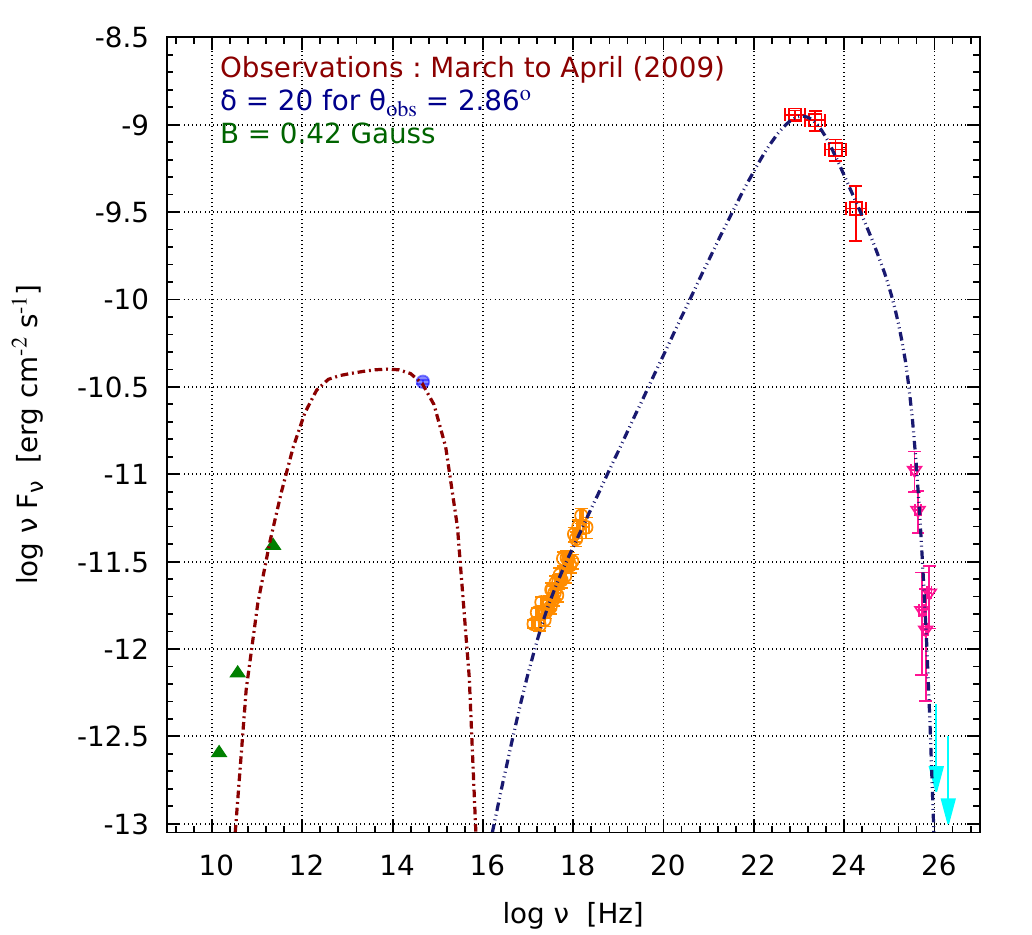} &
    \includegraphics[width=.48\textwidth]{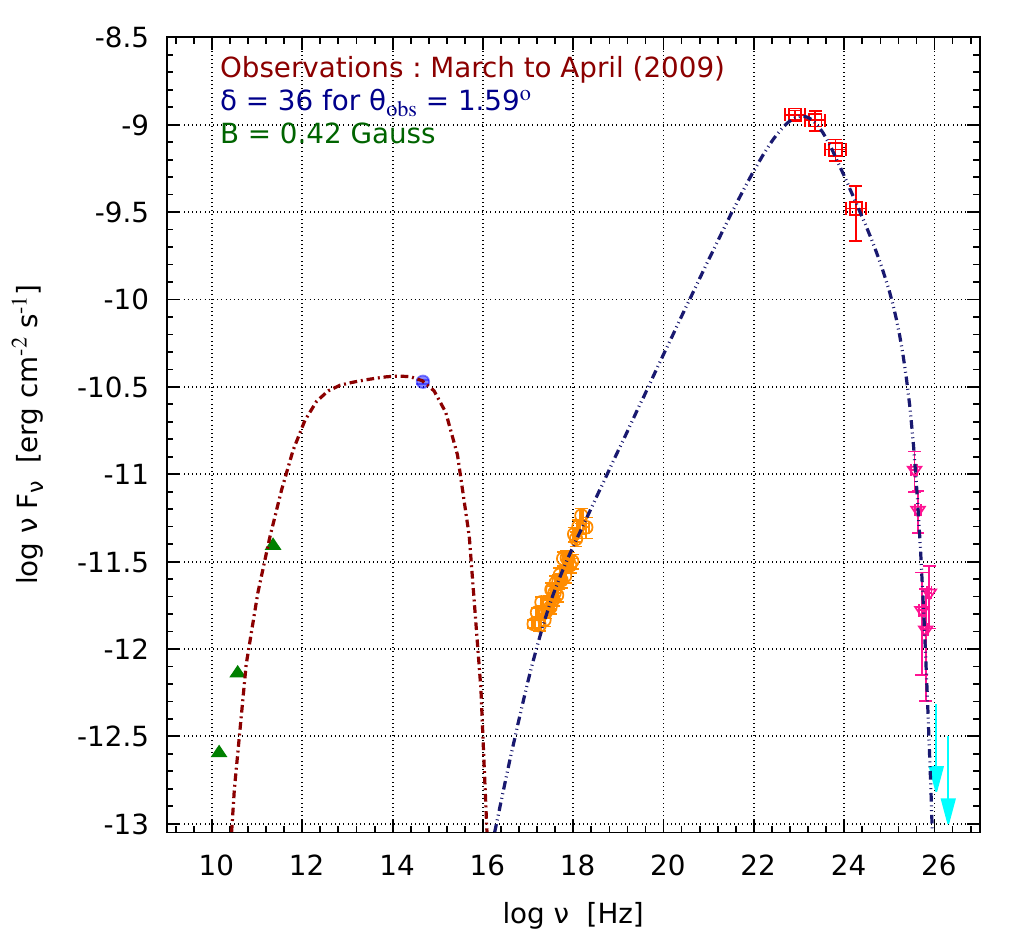}\\
    \includegraphics[width=.48\textwidth]{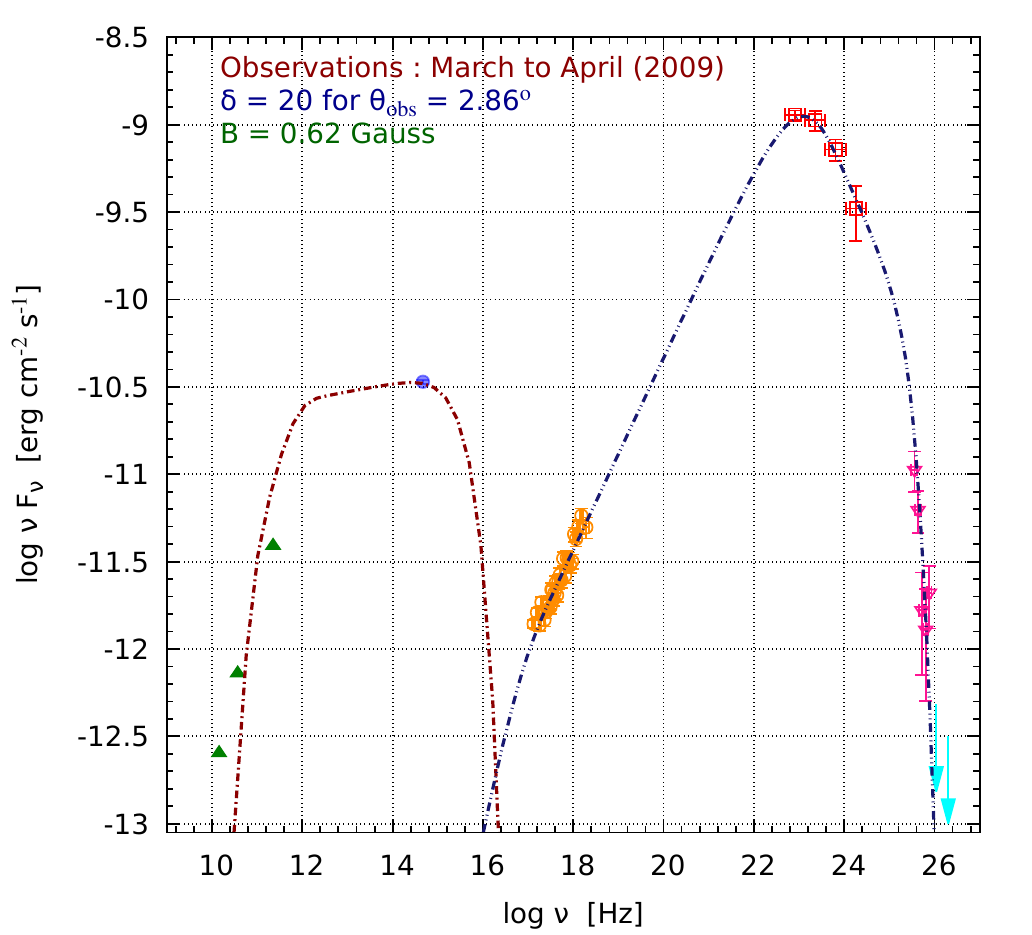} &
    \includegraphics[width=.48\textwidth]{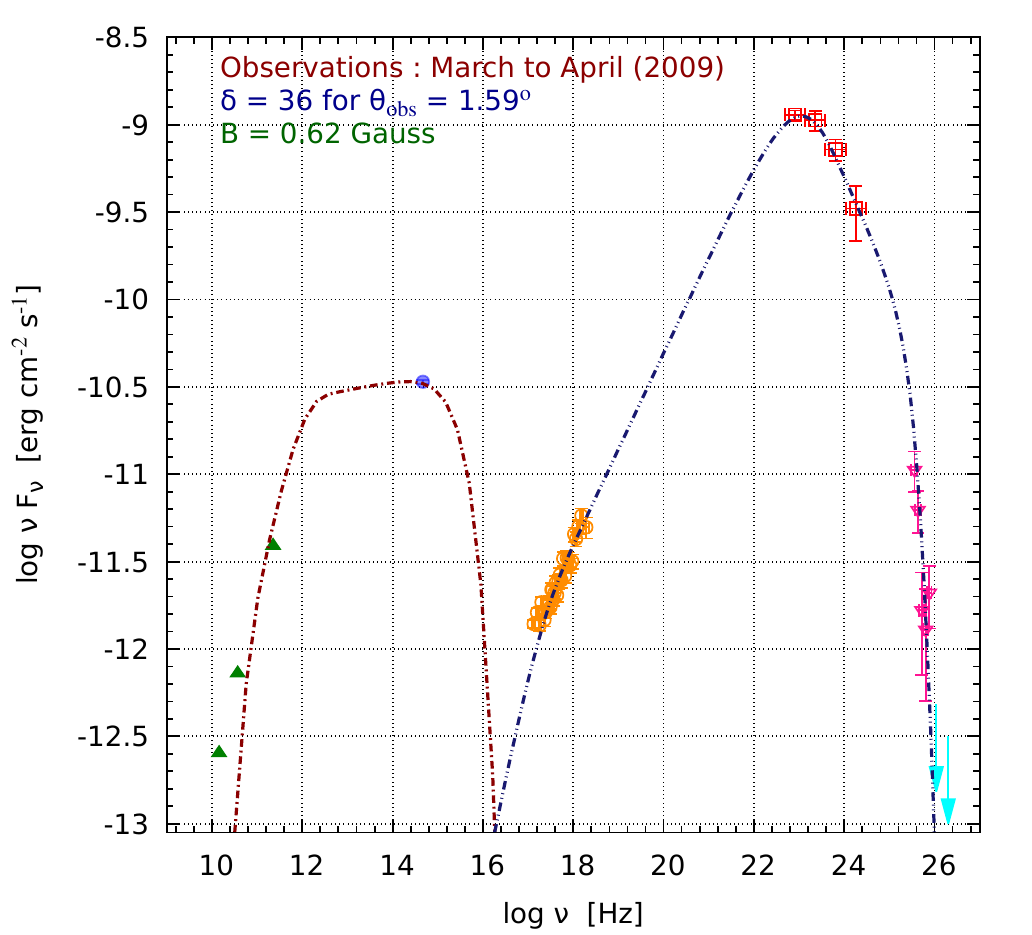}\\
  \end{tabular}  
  \caption{\textbf{Single zone synchrotron modeling of PKS 1510-089 assuming $\delta$ =20 $\&$ 36. Maroon single dot dashed line shows the electron synchrotron emission and dark blue double dot dashed line shows the proton synchrotron emission. Data points are shown from \cite{BK} --- i) Green filled upward pointing triangles---UMRAO, Mets\"ahovi and submillimeter array observations\protect\cite{Marscher}, ii) Blue translucent circle---ATOM telescope located on H.E.S.S site \cite{H.E.S.S.13}, iii) Orange hollow circles---XRT \cite{Evans} iv) Red hollow squares---\textit{Fermi}-LAT  \cite{Atwood}, v) Magenta hollow downward pointing triangles---\cite{H.E.S.S.13}, vi) Cyan downward arrows H.E.S.S upper limit. Top panel for \textit{B}=0.42 G \& bottom panel for \textit{B}=0.62 G.}}
    \label{Figure--1}
\end{figure*}

\begin{figure*}
\centering
  \begin{tabular}{@{}ccc@{}}
    \includegraphics[width=.48\textwidth]{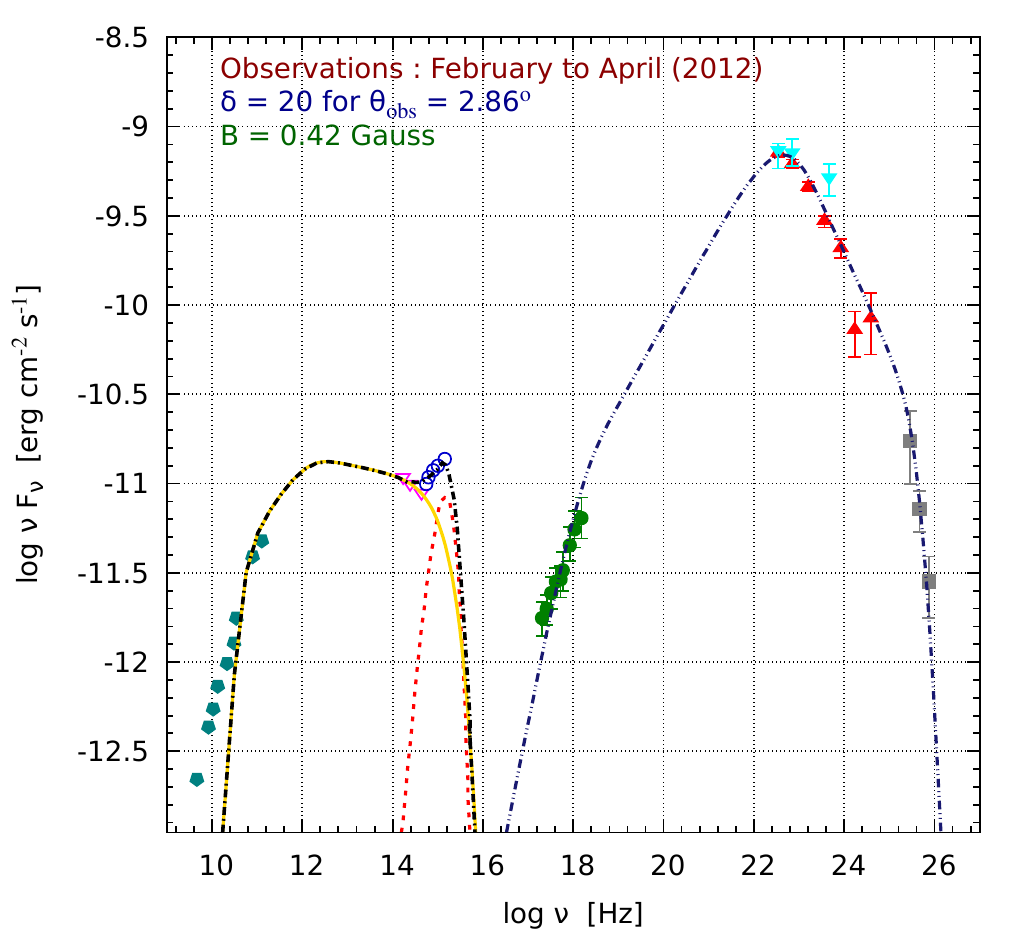} &
    \includegraphics[width=.48\textwidth]{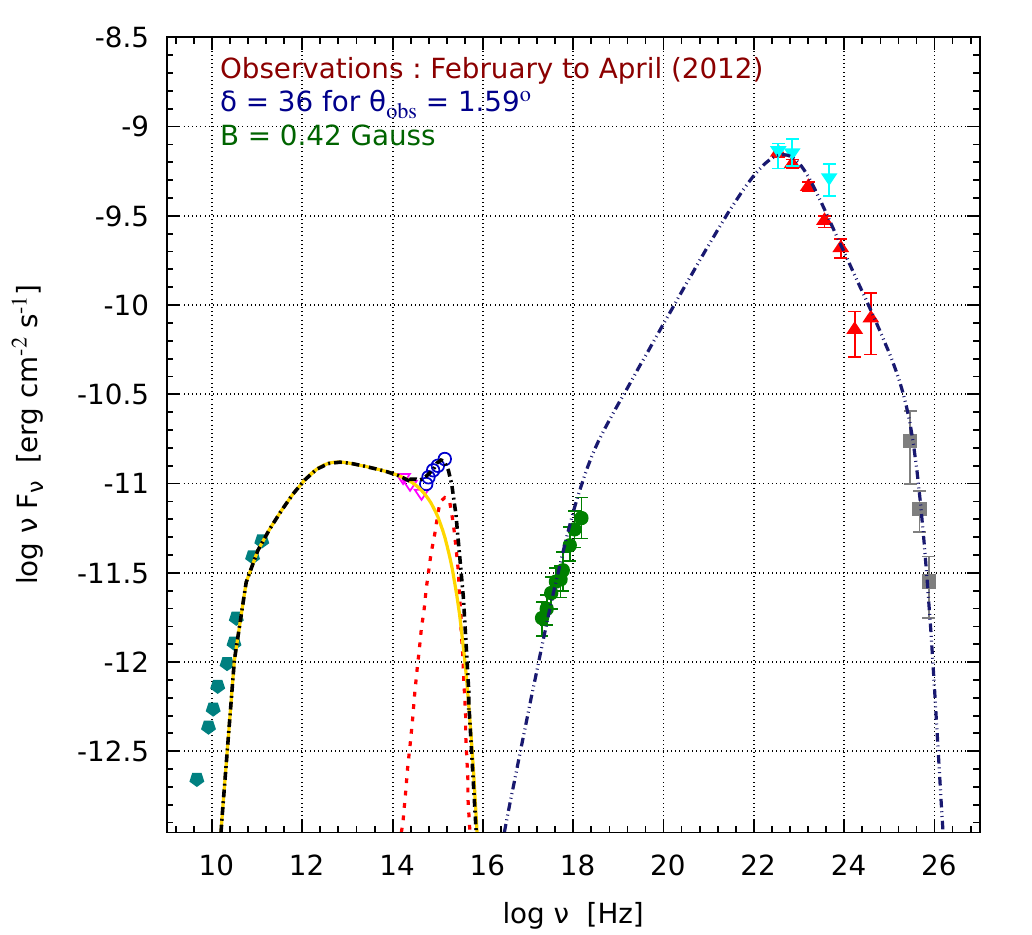}\\
    \includegraphics[width=.48\textwidth]{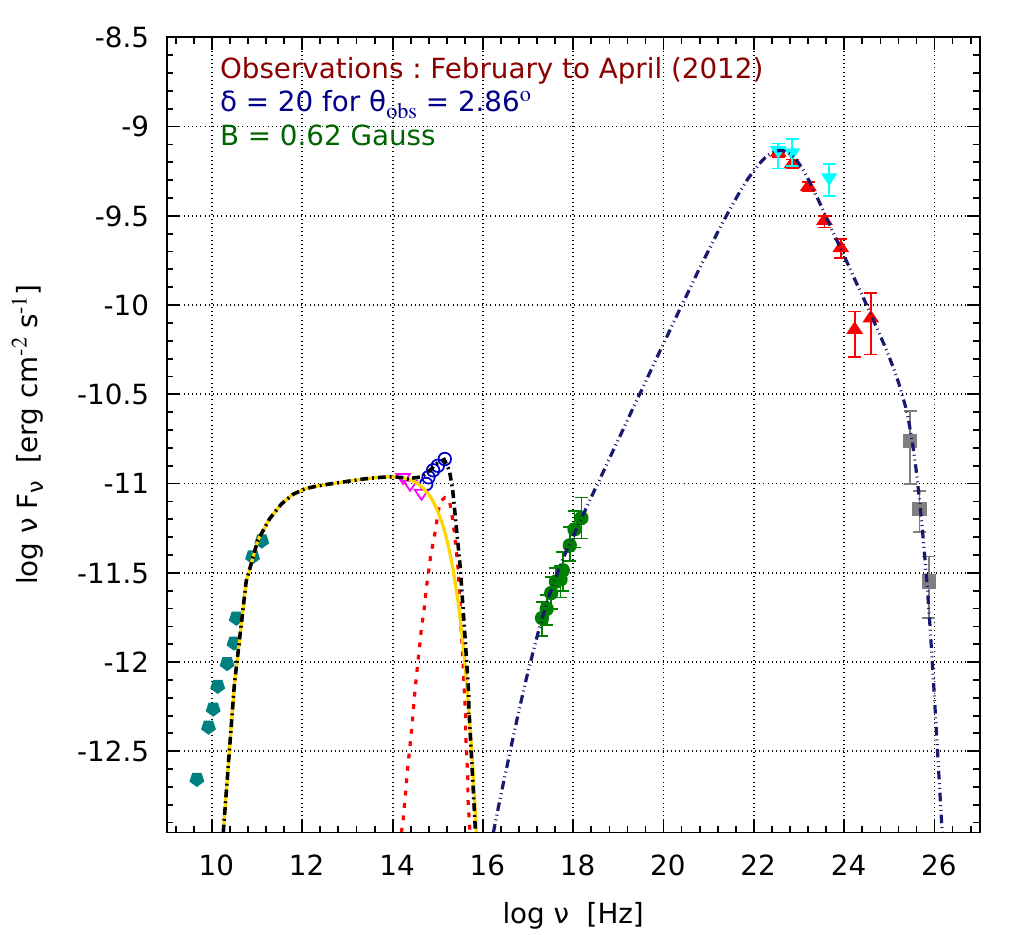} &
    \includegraphics[width=.48\textwidth]{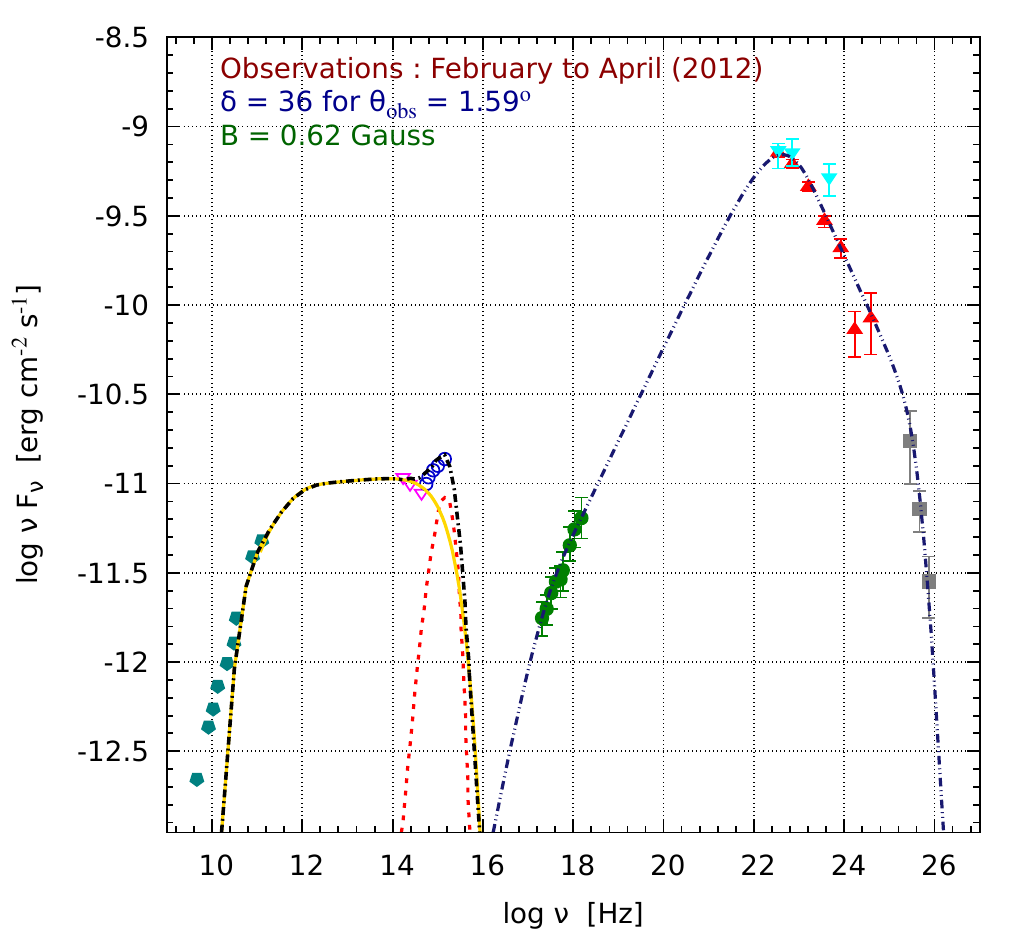}\\
  \end{tabular}  
    \caption{\textbf{Single zone synchrotron modeling of PKS 1510-089  assuming $\delta$ =20 $\&$ 36. The electron synchrotron emission is shown by the yellow continuous line and the emission from accretion disk is shown by the dashed red line. The combined emission spectra in the radio to optical--UV frequency is shown by the single dot dashed black line. The proton synchrotron emission is shown by the dark blue double dot dashed line. Data points are shown from \cite{Alek} --- i) Teal filled pentagons---F-GAMMA and Mets\"ahovi radio observations, ii) Magenta hollow downward pointing triangles---GASP-WEBT, iii) Blue hollow circles---\textit{Swift}-UVOT iv) Green filled circles---XRT, v) Red filled upward pointing triangles---\textit{Fermi}-LAT, vi) Cyan filled downward pointing triangles---AGILE-GRID(Flare-II) \cite{Lucarelli}, vii) Grey filled squares---\cite{Cortina}; \cite{Alek}. Top panel for \textit{B}=0.42 G \& bottom panel for \textit{B}=0.62 G.}}
    \label{Figure--2}
\end{figure*}

  As the VHE $\gamma$-rays are subjected to absorption by the extra-galactic background light(EBL) we have included the necessary correction to the $\gamma$-ray spectrum following the model of \cite{Franceschini}. The SEDs fitted by us are given in the next section (see  Figures\ref{Figure--1} \& \ref{Figure--2} ). The tables(Table--\ref{Table--1} $\&$ Table--\ref{Table--2}) show the specific values of the parameters used in our study. 

\begin{table*}

\centering\arraybackslash
\caption{}{\textbf{Parameters for fitting the multiwavelength data from the H.E.S.S flaring period (2009)}}
\scalebox{1.0}{
\begin{tabular}{|c|c|c|c|c|c|c|c|c|c|c|}
\hline
Synchrotron &$\Gamma$ &$\theta_{obs}$ &\textit{B} Gauss &\textit{R} cm &$E_{min}$ eV&$E_{max}$ eV &$E_{break}$ eV &$p_{1}$ &$p_{2}$ &$u_{particle}^{'}$ $\rm ergs/cm^{3}$ \\
\hline
Proton &20  &2.86$^{\circ}$  &0.42 &5.7$\times10^{17}$ &3.16$\times10^{15}$  &3.55$\times10^{19}$  &2.82$\times10^{18}$  &1.9  &4.15  &$5.8\times10^{-1}$  \\  
\hline
Electron &20 &2.86$^{\circ}$ &0.42 &5.7$\times10^{17}$ &9.33$\times10^{7}$ &3.16$\times10^{9}$ &1.18$\times10^{8}$ &1.9 &- &$5.8\times10^{-6}$ \\
\hline
Proton &36  &1.59$^{\circ}$  &0.42 &5.7$\times10^{17}$ &2.51$\times10^{15}$  &2.51$\times10^{19}$  &2.09$\times10^{18}$  &1.9  &4.15  &$7.4\times10^{-2}$  \\  
\hline
Electron &36 &1.59$^{\circ}$ &0.42 &5.7$\times10^{17}$ &5.81$\times10^{7}$ &3.16$\times10^{9}$ &1.18$\times10^{8}$ &1.9 &- &$7\times10^{-7}$ \\
\hline
Proton &20  &2.86$^{\circ}$  &0.62 &5.7$\times10^{17}$ &1.66$\times10^{15}$  &2.82$\times10^{19}$  &2.39$\times10^{18}$  &1.9  &4.15  &$3.25\times10^{-1}$  \\  
\hline
Electron &20 &2.86$^{\circ}$ &0.62 &5.7$\times10^{17}$ &5.37$\times10^{7}$ &4.73$\times10^{9}$ &5.42$\times10^{7}$ &1.9 &- &$3.8\times10^{-6}$ \\
\hline
Proton &36  &1.59$^{\circ}$  &0.62 &5.7$\times10^{17}$ &2.18$\times10^{15}$  &2.18$\times10^{19}$  &1.69$\times10^{18}$  &1.9  &4.15  &$41.75\times10^{-3}$  \\  
\hline
Electron &36 &1.59$^{\circ}$ &0.62 &5.7$\times10^{17}$ &5.37$\times10^{7}$ &3.16$\times10^{9}$ &5.42$\times10^{7}$ &1.9 &-&$3.8\times10^{-7}$ \\
\hline
\end{tabular}}
\label{Table--1}
\end{table*}
 \begin{table*}

\centering\arraybackslash
\caption{}{\textbf{Parameters for fitting the multiwavelength data from the MAGIC flaring period (2012)}}
\scalebox{1.0}{
\begin{tabular}{|c|c|c|c|c|c|c|c|c|c|c|}
\hline
Synchrotron &$\Gamma$ &$\theta_{obs}$ &\textit{B} Gauss &\textit{R} cm &$E_{min}$ eV&$E_{max}$ eV &$E_{break}$ eV &$p_{1}$ &$p_{2}$ &$u_{particle}^{'}$ $\rm ergs/cm^{3}$\\
\hline
Proton &20  &2.86$^{\circ}$  &0.42 &5.7$\times10^{17}$ &9.66$\times10^{15}$  &7.41$\times10^{19}$  &1.82$\times10^{18}$  &2.1  &4.1  &$7.35\times10^{-1}$  \\  
\hline
Electron &20 &2.86$^{\circ}$ &0.42 &5.7$\times10^{17}$ &$10^{6}$ &3.63$\times10^{9}$ &1.18$\times10^{8}$ &2.1 &- &$1.35\times10^{-5}$ \\
\hline
Proton &36  &1.59$^{\circ}$  &0.42 &5.7$\times10^{17}$ &6.31$\times10^{15}$  &7.41$\times10^{19}$  &1.35$\times10^{18}$  &2.1  &4.1  &$9.7\times10^{-2}$  \\  
\hline
Electron &36 &1.59$^{\circ}$ &0.42 &5.7$\times10^{17}$ &6.31$\times10^{5}$ &2.82$\times10^{9}$ &1.18$\times10^{8}$ &2.1 &- &$1.42\times10^{-6}$ \\
\hline
Proton &20  &2.86$^{\circ}$  &0.62 &5.7$\times10^{17}$ &2.82$\times10^{15}$  &7.41$\times10^{19}$  &1.32$\times10^{18}$  &1.92  &4.1  &$3.65\times10^{-1}$  \\  
\hline
Electron &20 &2.86$^{\circ}$ &0.62 &5.7$\times10^{17}$ &6.31$\times10^{5}$ &2.99$\times10^{9}$ &5.42$\times10^{7}$ &1.92 &- &$5.8\times10^{-6}$ \\
\hline
Proton &36  &1.59$^{\circ}$  &0.62 &5.7$\times10^{17}$ &2.24$\times10^{15}$  &7.41$\times10^{19}$  &1.05$\times10^{18}$  &1.95  &4.1  &$4.5\times10^{-2}$  \\  
\hline
Electron &36 &1.59$^{\circ}$ &0.62 &5.7$\times10^{17}$ &6.31$\times10^{5}$ &2.51$\times10^{9}$ &5.42$\times10^{7}$ &1.95 &- &$6.2\times10^{-7}$ \\
\hline
\end{tabular}}
\label{Table--2}
\end{table*}

    \begin{table*}
\centering
\caption{}{\textbf{ Jet power in lepto-hadronic models for the flaring states of PKS 1510-089}}    
\begin{center}
\scalebox{1.0}{
\begin{tabular}{ |c|c|c|c|c|c|c|c| } 
 \hline
 Obs &$B$ Gauss & \textit{R} cm&$\Gamma$  &$P_{jet}^{electron}$ ergs/sec &$P_{jet}^{proton}$ ergs/sec & $P_{jet}^{magnetic}$ ergs/sec & $P_{jet}^{total}$ ergs/sec \\ 
 \hline
  & 0.42 & &20 &$7.08\times10^{43}$ &$7.08\times10^{48}$&$8.57\times10^{46}$ &$7.17\times10^{48}$\\ 
 2009 & 0.42 &$5.7\times10^{17}$ &36 &$2.77\times10^{43}$ &$2.93\times10^{48}$&$2.78\times10^{47}$ &$3.20\times10^{48}$\\  
   & 0.62 & &20 &$3.66\times10^{43}$ &$3.97\times10^{48}$&$1.87\times10^{47}$ &$4.15\times10^{48}$\\ 
  & 0.62 & &36 &$1.42\times10^{43}$ &$1.65\times10^{48}$&$6.05\times10^{47}$ &$2.26\times10^{48}$\\  
 \hline
   & 0.42 & &20 &$1.65\times10^{44}$ &$8.97\times10^{48}$&$8.57\times10^{46}$ &$9.06\times10^{48}$\\ 
 2012 & 0.42 &$5.7\times10^{17}$ &36 &$5.62\times10^{43}$ &$3.84\times10^{48}$&$2.78\times10^{47}$ &$4.11\times10^{48}$\\  
   & 0.62 & &20 &$7.08\times10^{43}$ &$4.46\times10^{48}$&$1.87\times10^{47}$ &$4.64\times10^{48}$\\ 
  & 0.62 & &36 &$2.45\times10^{43}$ &$1.78\times10^{48}$&$6.05\times10^{47}$ &$2.38\times10^{48}$\\  
 \hline
\end{tabular}}
\end{center}
\label{Table--3}
\end{table*}         
\section{Results and Discussions}
       The spectral energy distributions of PKS 1510-089 during the epochs [March--April(2009); February--April(2012)]\footnote{results and parameter details are presented in this chronological order unless otherwise mentioned}, when it was active in the VHE regime, have been  modeled using a single zone lepto-hadronic model. The size of the emission region assumed in our model is \textit{R}=5.7$\times10^{17}$ cm and magnetic field \textit{B}=0.42 \& 0.62 Gauss. 
      The injection spectrum of relativistic electrons is assumed to be a simple power law in our model. 
      Due to synchrotron cooling a break appears in the propagated spectrum of electrons.
      The spectral indices of the propagated electron spectrum are $p_1$ and $p_1+1$ below and above the break energy respectively.
      The break energies in the propagated spectrum of electrons evaluated using eqn(\ref{eqn_2}) for the magnetic fields 0.42 G and 0.62 G are 1.18$\times10^{8}$ eV $\&$ 5.42$\times10^{7}$ eV respectively, assuming the same value of \textit{R}  in both cases. For modeling the lower energy bump as observed during the flaring epoch of 2009 we require the electron population to be in the energy range $10^{7}${eV} $<E^{electron}<10^{10}\rm{eV}$  whereas for the 2012 observations the electrons should be in this energy range $0.6\times10^{6}${eV} $<E^{electron}<10^{10}\rm{eV}$.  
      \par
      In our model the relativistic protons lose energy by synchrotron emission in the magnetic field of the jet. It is worth mentioning in this context that the relativistic protons may radiate energy by both proton synchrotron mechanism and/or photo-meson production.  The synchrotron emission by the relativistic electrons may provide the seed photons for $p\,\gamma$ interactions.  Synchrotron self Compton emission (SSC)  is  very low in our model as the size of the emission region is large.
      The threshold energy condition for the production of photo mesons  in the jet frame is
      \begin{equation}
      E^{proton} \, \epsilon \geq 0.14 \, {\rm GeV^2}
      \label{eqn_3}
      \end{equation} 
      where $\epsilon$ is the energy of the photons emitted in synchrotron cooling of electrons.
With the elasticity-weighed $p\gamma$ interaction cross section $\langle\sigma_{p\gamma}f\rangle$ and density of seed photons $n_{ph}(\epsilon)$ having energy $\epsilon$ to 
$\epsilon_{max}$ (which is the maximum energy of the synchrotron photons in the jet frame corresponding to the maximum energy of electrons $E_{max}^{electron}$),  the timescale of $p\,\gamma$  interactions -- $t_{p\gamma}$  is      
\begin{equation}
t_{p\gamma}=\dfrac{1}{c \, \langle\sigma_{p\gamma}f\rangle \, n_{ph}(\epsilon)} \, \rm  secs
\label{eqn_4}
\end{equation}
    It  is found to be greater than the synchrotron cooling time of the protons for the parameter values used in our work.  The synchrotron loss time scale of the protons having maximum energy $7.41\times 10^{19}$ eV in a magnetic field of 0.62 Gauss is $1.55\times 10^{8}$ secs which is longer than the duration of the flares $\Delta t_{jet}=\Delta t_{obs} \delta = 4.36\times 10^7$ secs (where $\Delta t_{obs}=$ 14 days, $\delta=36$ ) and the light crossing time $\frac{R}{c} = 1.91\times 10^{7}$ secs. The corresponding value of $t_{p\gamma}$ for the above values of \textit{B}, $\delta$ and $E_{max}^{proton}$ (corresponding to the modeling of 2012 flaring episode) is $1.98\times 10^{10}$ secs. In our study the parameter values are chosen such that the condition $t_{p\gamma} > t^{proton}_{synch} > R/c$  is always satisfied. 
      As the cooling time scales of the protons are long compared to the duration of the flares,  the propagated spectrum of protons remains essentially  same as their injected spectrum. Although the spectral index \(p_1\)$\sim 2$  may come from shock acceleration in the jet a sharp attenuation  in the injected spectrum of protons beyond the break is difficult to explain with conventional acceleration mechanisms. 
      
      \par 
      
      The observed data in the X-ray to VHE $\gamma$-ray frequency range give  the second energy bump in the SEDs of PKS 1510-089. Our results indicate that synchrotron emission of a population of relativistic protons in the energy range $10^{15}\rm{eV}$ $<E^{proton}<10^{20}\rm{eV}$ can reproduce the second  bump in both the SEDs satisfactorily. The  maximum energy of protons used  in our model  is $7.41\times10^{19}$ eV for which the Larmor radii for the magnetic fields  0.42 Gauss and 0.62 Gauss  are $5.29\times10^{17}$ cm and  $3.59\times10^{17}$ cm respectively. These values are less than the radius of our emission region  $R=5.7\times10^{17}$ cm. The fits according to our model for the SED of  2009 flare is shown in Figure--\ref{Figure--1} and for the 2012  flare in Figure--\ref{Figure--2}. All the model parameters of our lepto-hadronic framework are tabulated in Table--\ref{Table--1} and Table--\ref{Table--2}. Our figures show that  a single-zone lepto-hadronic model  can provide  good description of the overall broadband spectrum  during  flaring states of PKS 1510-089. 
           
    \par  
      Earlier  in \cite{BK} the authors have used a broken power law distribution of injected electrons having spectral indices \(p_1\) = 1.2 \& \(p_2\) = 3.4 below and above the break energy respectively to describe the flare of 2009 with the external Compton model. It is difficult to explain the spectral indices of 1.2, 3.4 and the break in their injected electron spectrum with conventional acceleration mechanisms. They used the BLAZAR code in their work to obtain the propagated electron spectrum and subsequently their multi-wavelength emission spectrum. 
    \par  
      The authors of \cite{Alek} have assumed that the emission region is filled with relativistic electrons having a broken power law spectrum to model the flare of 2012 with the external Compton model. They have used spectral indices 1.9 and 3.85 below and above the break energy for their electron spectrum while considering  external Compton emission with the seed photons of infrared-torus. Similar values are also assumed in their model for external Compton emission with the seed photons from  spine and sheath regions.
 Our injected spectrum of electrons is a single power law with $p_1\sim 2$ in all cases while modeling the low energy bumps of the SEDs.   As mentioned earlier break appears in the propagated spectrum of the electrons in our work due to synchrotron cooling. Beyond the break energy the spectral index  $p_1+1$  is physically consistent in our model.

      \par There is only a single datapoint in the optical--UV frequency range in the SED of 2009 and the so called ``big blue bump" which is attributed to the blackbody radiation from the accretion disk is not present. Hence while modeling the lower energy bump of the 2009 SED we have solely relied on the synchrotron emission from electrons to account for both the radio and optical data. On the other hand the multiwavelength data accumulated during the 2012 flare shows a bump in the optical--UV range which clearly indicates thermal emission from the accretion disk. We have added  the synchrotron photon flux from electrons to the thermal emission  from the accretion disk assuming a disk luminosity of $6.7\times 10^{45}$ ergs/sec to describe the observed data in the optical--UV range satisfactorily in this case. We note that in \cite{Alek} the authors have also assumed the same value for the disk luminosity while fitting the SED with their model.

     \subsection{Jet power in lepto-hadronic model for PKS 1510-089}
    The power of relativistic jet in the observer's frame is evaluated with the expression given below
    \begin{equation}
    P_{jet}=\pi R^{2} \Gamma^{2} c (u_{electron}^{'}+u_{proton}^{'}+u_{B}^{'})
    \label{eqn_5}
    \end{equation}
where $u_{electron}^{'}$, $u_{proton}^{'}$ and $u_{B}^{'}$ are the energy densities of the electrons, protons and the magnetic field in the jet frame. The total jet power required in all our models along with the individual contributions from electrons, protons and the magnetic field are tabulated above in Table--(\ref{Table--3}).

 As can be seen the jet power in all our models are of the order of $10^{48}$ ergs/sec. The minimum value of jet power for the 2009 observations in our model is $2.26\times 10^{48}$ ergs/sec which was achieved for ${B} = 0.62$ G \& $\delta$ = 36. Whereas for the 2012 observations the minimum jet power was $2.38\times 10^{48}$ ergs/sec for the same values of $B$ \& $\delta$. While modeling the HE peak of PKS 1510-089 (excluding VHE data) the authors of \cite{Bo13} have reported the contribution of protons alone to the jet power as 2.5$\times10^{48}$ergs/sec and as they have assumed equipartition between the protons and magnetic field the total jet power in their model is $\sim$5$\times10^{48}$ergs/sec. Although our results are in agreement with the results obtained in \cite{Bo13}, it is evident that the jet power is very high especially in comparison with the Eddington luminosity\footnote{$L_{Edd}$ is calculated from the estimate of black hole mass given in \cite{Abdo10}}($L_{Edd}$=6.86$\times10^{46}$ ergs/sec) and the disk luminosity ($10^{46}$ergs/sec---\cite{Pucella}) . However according to \cite{Tchekhovskoy} general relativistic, magnetohydrodynamic numerical simulations show that the outflowing energy may surpass the energy flowing in to the central black hole by a considerable amount. Moreover according to \cite{Ghisellini}, for the majority of blazar sources studied by them $\frac{P_{jet}}{L_{disk}}$=10, in some cases this ratio even goes onto 100. These studies  indicate that the jet power values might indeed reach magnitudes of $10^{47}$---$10^{48}$ ergs/sec for PKS 1510-089 during its high  states. Also in \cite{Abdo10} the authors have reported one of the highest $\gamma$--ray isotropic luminosity of 2$\times10^{48}$ ergs/sec on 2009 March 26 for PKS 1510-089. Moreover, a luminosity value of $\sim$ 2$\times 10^{49}$ ergs/sec was reported corresponding to the peak flux on $19^{th}$ October 2011 \cite{Foschini}. The isotropic luminosity required in our model is of the order of  $\sim 10^{50}$--$10^{51}$ ergs/sec. Hence the possibility of attaining super-Eddington luminosity during flares cannot be ruled out.
 
     \par We note that the higher energy bump in the SEDs could have been modeled for considerably higher magnetic fields but in that case it is not possible to account for the radio frequency data as consistently as done in the present work. Also a higher magnetic field would mean that the contribution of $P_{jet}^{magnetic}$ to the total jet power would increase considerably without resulting in any sufficient decrease in $P_{jet}^{proton}$ contribution. 
     On the other hand if we decrease the magnetic field  below 0.42 G then the jet luminosity from protons  $P_{jet}^{proton}$ tends to exceed $10^{49}$ ergs/sec.
                
\section{Concluding Remarks}

We have discussed about a single zone lepto-hadronic model to fit the SEDs of PKS 1510-089 during its high states in 2009 and 2012. 
As our choice of $\delta$ values is guided by the fact that $\theta_{obs}$ can lie within the range 1.4$^{\circ}$---3$^{\circ}$ (as mentioned earlier in section--(\ref{section_2})) we can safely conclude that a lepto-hadronic model based explanation for the flaring states of PKS 1510-089 is applicable for a very wide range $1.59^{\circ}\leqslant \theta_{obs} \leqslant 2.86^{\circ} $. 
Due to inefficient cooling of protons we need very high jet power in protons to explain the higher energy bumps in the SEDs of flares.
Although the jet power in our model is of the order of $10^{48}$ ergs/sec (which corresponds to isotropic luminosity of the order of $10^{50}$--$10^{51}$ ergs/sec), such high luminosities might be attained during blazar flares [see \cite{Abdo10} $\&$ \cite{Foschini}]. Earlier studies also indicated the need of such high luminosities in lepto-hadronic model (see \cite{Bo13}). 
\par The dimension of the emission region in our model is consistent with the observed week scale variability in the HE and radio frequency range. The hour scale variability reported by \cite{Saito} and \cite{Foschini} may be attributed to smaller `cells' inside the larger emission region. As discussed in \cite{Marscher14} the progression of turbulent plasma at relativistic velocities may give rise to the shorter timescale variabilities. Fluctuations in magnetic field and particle density may also play a significant role behind the rapid variability or ``flickering". Although no variability in the VHE regime was observed during the flaring episodes under discussion in our paper, it is probable that owing to reasons stated above the source may exhibit rapid variability in the VHE regime as well.

\section*{Acknowledgements}
We are indebted to  Anna Barnacka for providing us the data used in \cite{BK}.

\end{document}